\documentclass[preprint,aps,epsfig]{revtex4}
\usepackage{amsmath}
\usepackage{amssymb}
\usepackage{graphicx}
\usepackage{dcolumn}
\usepackage{bm}
\newcommand{\beq}{\begin{eqnarray}}
\newcommand{\eeq}{\end{eqnarray}}
\newcommand{\bea}{\begin{eqnarray}}
\newcommand{\eea}{\end{eqnarray}}

\begin{document}

\title{\textbf{Hadronic current correlation functions at \\ finite temperature
in the NJL model}}
\author{R.L.S.Farias}
\email{ricardof@ift.unesp.br}
\affiliation{Instituto de F\'\i sica Te\'orica, Universidade Estadual Paulista \\
Rua Pamplona, 145, 01405-900, S\~ao Paulo, SP - Brazil}
\author{G. Krein}
\email{gkrein@ift.unesp.br}
\affiliation{Instituto de F\'\i sica Te\'orica, Universidade Estadual Paulista \\
Rua Pamplona, 145, 01405-900, S\~ao Paulo, SP - Brazil}

\author{O.A.Battistel}
\email{orimar@ccne.ufsm.br}
\affiliation{Departamento de F\'{\i}sica, Universidade Federal de Santa Maria \\
  POBox 5093, 97119-900 Santa Maria, RS - Brazil}

\vspace{2 cm}
\begin{abstract}
Recently there have been suggestions that for  a
proper description of hadronic matter and hadronic correlation
functions within the NJL model at finite density/temperature the
parameters of the model should be taken density/temperature
dependent. Here we show that qualitatively similar results can be
obtained using a cutoff-independent regularization of the NJL model.
In this regularization scheme one can express the divergent parts at
finite density/temperature of the amplitudes in terms of their
counterparts in vacuum.
\end{abstract}

\maketitle

Recently there have been suggestions that for a proper description
of hadronic matter and hadronic correlation functions within the
Nambu--Jona-Lasinio (NJL) model at finite densities and
temperatures the parameters of the model should be taken
temperature and density dependent. In a study of the color flavor
locked phase of QCD within the NJL model Casalbuoni et
al.~\cite{casalbuoni} have shown that in order to get sensible
results the ultraviolet cutoff of the model should increase with
density. Similarly, Shakin et al. \cite{shakin1,shakin2} in
studies of hadronic correlation functions found that the NJL model
may have a broader range of application than previously considered
if one allows a significant temperature dependence of the
parameters of the model, in particular of the cutoff.

The aim of the present communication is to show that qualitatively
similar results of Refs.~\cite{shakin1} and \cite{shakin2} can be
obtained using a cutoff-independent regularization of the NJL
model~\cite{BN,BK}. In this novel regularization scheme, one
assumes an implicit regularization of the divergent integrals and
by making use the scaling properties of these integrals, one can
express the divergent parts of the amplitudes at finite
temperature/density in terms of their counterparts at zero
temperature/density. At one loop there are only two divergent
integrals and by fitting these to physical quantities at zero
temperature and density, no cutoff needs to be introduced. One
consequence of this is that all momentum integrals over the
Fermi-Dirac distributions can be extended to infinity, and need
not be cutoff at some large mass scale. Here we apply this
regularization scheme to study finite temperature hadronic current
correlation functions~\cite{us}.

At the level of the mean field (or one loop) approximation, there
appear two divergent integrals, one quadratically divergent
($I_{q}$) and one logaritmically divergent ($I_{l}$):
\begin{eqnarray}
I_{q}(M^2) = \int_{reg}\frac{d^4k}{(2\pi)^4}\frac{1}{(k^2 -
M^2)}\hspace{1.25cm} I_{l}(M^2) = \int_{reg} \frac{d^4k}{(2\pi)^4}
\frac{1}{(k^2 -M^2)^2}
\end{eqnarray}
It is not difficult to show that these integrals satisfy the
following scaling properties~\cite{BN},
\begin{eqnarray}
I_{q}({M}^2) &=& I_{q}(\mu^2) + ({M}^2 - \mu^2)I_{l}(\mu^2) +
\frac{i}{(4\pi)^2}\left[ {M}^2 - \mu^2 - {M}^2 \,
{\log}({M}^2/\mu^2) \right] \label{scal-Iq}\\
I_{l}(M^2) &=& I_{l}(\mu^2) - \frac{i}{(4\pi)^2}{\log} (M^2/\mu^2)
\label{scal-Il}
\end{eqnarray}
where $\mu$ is an arbitrary mass scale. In general, when dealing
with finite density/temperature, the mass scale that appears in
$I_q$ and $I_l$ is density/temperature dependent. The obvious use
one can make from the scaling relations of Eqs.~(\ref{scal-Iq})
and (\ref{scal-Il}) is to express the divergent integrals at
finite temperature in terms of the zero temperature integrals.
Moreover, since the finite parts are integrated over all momenta,
temperature effects coming from the Fermi-Dirac distribution
function are not cutoff at some finite momentum as in cutoff
regularization schemes. In this way, there is no need to use
arbitrary temperature dependencies on the cutoffs~\cite{{shakin1}}
in order to extend the range of momentum integrals. In order to
illustrate these properties, we consider the pseudoscalar
correlation function at high temperatures. In particular, we will
show how the regularization scheme naturally isolates the finite,
temperature-dependent contributions from purely divergent, vacuum
effects, and compare with the corresponding results using a cutoff
scheme.

The pseudoscalar correlator $C_P(P^2,T)$ is the Fourier transform
of the thermal average of the time-ordered product of pseudoscalar
currents $j_5 = \bar\psi\gamma_5 \tau^i\psi$. The imaginary part
of $C_P(P^2,T)$ can be written as~\cite{shakin1}
\begin{eqnarray}
{\rm Im}\,C_{P}(P^2,T) = \frac{{\rm Im}\,\Pi_{P}(P^2,T)} {[1 - 2
\, G\, {\rm Re}\Pi_P(P^2,T)]^2 + [2 \, G \,{\rm Im}
\Pi_P(P^2,T)]^2}
\end{eqnarray}
where $\Pi_P(P^2,T)$ is the pseudoscalar polarization function. In
the frame $\vec P=0$, the imaginary part of $\Pi_P(P^2,T)$ is
given by
\begin{eqnarray}
{\rm Im} \, \Pi_{P}(\omega^2,T) = \frac{N_c \,\omega}{4\pi}
(\omega^2 - 4{M^*}^2)^{1/2} [1 - 2 n(\omega/2)]
\end{eqnarray}
where $M^*$ is the temperature-dependent constituent quark mass,
and $n$ is the Fermi-Dirac distribution function. The real part of
$\Pi_P(\omega^2,T)$ can be obtained by the dispersion integral,
which turns out to be ultraviolet divergent. As announced
previously, one can express the divergent part of the amplitude in
terms of $I_{q}$ and $I_{l}$:
\begin{eqnarray}
{\rm Re} \, \Pi_P(\omega^2,T) = \frac{N_c}{4\pi^2}{\cal P}
\int_{4{M^*}^2}^\infty  \frac{ds[\omega^4 -
2s^2n(\sqrt{s}/2)]}{s^{3/2}(s - 4{M^*}^2)^{1/2}} \,  - 4 N_c
\omega^2 \left[iI_{l}(4{M^*}^2) -
\frac{iI_{q}(4{M^*}^2)}{2\omega^2}\right] \nonumber
\end{eqnarray}
The principal value integral is finite. Using the scaling
relations above, one can now re-express $I_{l}(4{M^*}^2)$ and
$I_{q}(4{M^*}^2)$ in terms of $I_{l}(M^2)$ and $I_{q}(M^2)$. These
last two integrals are vacuum quantities and can be fitted to
chiral parameters, like $\langle\bar\psi\psi\rangle$ and
$f_{\pi}$.

In Figure~1 we plot the integrals
\begin{eqnarray}
F_{cut}(\omega^2,T) &=& {\cal P}\int_{4{M^*}^2}^\infty ds\;
\frac{\sqrt{s\; (s-4{M^*}^2)}}{s - \omega^2}\, n(\sqrt{s}/2) \,
e^{-(s-4M^{*2})/4\alpha^2(T)}
\end{eqnarray}
and
\begin{eqnarray}
F_{imp}(\omega^2,T) &=& {\cal P}\int_{4{M^*}^2}^\infty ds\,
\frac{[\omega^4 - 2s^2n(\sqrt{s}/2)]}{s^{3/2}(s - 4{M^*}^2)^{1/2}}
\end{eqnarray}
where $\alpha(T)$ is the temperature-dependent cutoff. These
functions are precisely the parts of the integrals that depend on
the Fermi-Dirac distribution. While in $F_{cut}$ the high energy
region is cutoff by the form factor, $F_{imp}$ is not - but its
ultraviolet behavior is softened by the subtractions done in order
to extract the infinite contributions. We use $T = 1.5~T_c$, with
$T_c = 150$~MeV, for which $M^* = 10$~MeV. The Figure shows that
the effect of a temperature dependent cutoff is very well captured
by the implicit regularization.

\vspace{0.5cm}
\begin{figure}[h]
\includegraphics[height=.38\textheight]{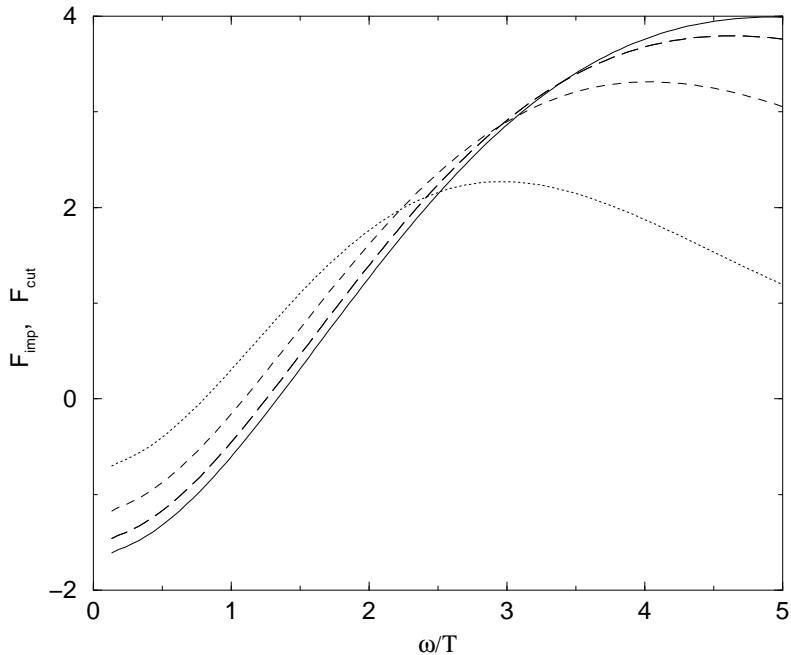}
\caption{The functions $F_{imp}(\omega^2,T)$ (solid line) and
$F_{cut}(\omega^2,T)$ for different values of the cutoff: $\alpha
= 0.5$~GeV (\mbox{dotted)}, $\alpha = 1.0$~GeV (\mbox{dashed}),
$\alpha = 2.0$~GeV (\mbox{long-dashed}).}
\end{figure}

\section{Acknowledgments} Work partially financed by CAPES, CNPq and
Fapesp.

\end{document}